# Inexpensive hetero-core spliced fiber optic setup for assessing strain

R. Biswas*

Applied Optics and Photonics Laboratory, Department of Physics, Tezpur University, Tezpur-784028, Assam

email: rajib@tezu.ernet.in

Abstract: We report here a novel experimental scheme for the measurement of strain that utilizes a simple intensity based interrogation system, based on optical fibers. Corresponding to splicing of a multimode fiber between two single mode fibers, the transmittance was measured by subjecting them to strain at multi points. The results corresponding to those multi points are found to mimic MMI interfere considerably. The set-up has the potential to be applied in crack propagation analysis.

1. Introduction

This recent decade has seen a considerable growth in the domain of fiber optics sensing. In sensing design, hetero core splicing has become a very robust too for measurement of parameters. To be precise, the underlying working principle is multimode interference. Based on these, several research workers reported many novel optical devices, e.g. a displacement sensor, a fiber lens, a refractometer sensor, an edge filter for wavelength measurements etc [1-26].

Of late, strain measurement has become quite important. Due to its allied relations with other physical parameters, experimentalists emphasize on precise measurement of this. The estimated value will then help them to assess other allied terms. However, level of strain measurement may be lower or higher depending on target. For instance, strain measurement of microstrain level bears an important role in checking integrity of structures, such as in civil bridges, aircraft materials etc. In order to assess this, several implementations are there in terms of sensing. Among the designs or schemes, experimentalists prefer fiber sensor to others. Thanks to the immunity to electromagnetic interference [3-7]. There are several adaptations of fiber sensing schemes towards measurement of this parameter [9-11]. Fiber Bragg Grating needs special mention. Although it is a robust system, the use of interrogator brings complexity in the adaptation. Besides, fabrication part is also tedious [12-13]. These lacunae motivate researchers to explore other fiber optic sensing adaptations. As a result, the idea of splicing hetero-core fibers rapidly took shape. Now, there is a copious amount of literatures highlighting use of these schemes. In this direction, single mode-multimode-single mode or multimode-single mode-multimode has been extensively used in assessing different parameters [14-18]. SMS fiber configurations have been reported to be used in strain measurement. However, measuring distribution of strain across a certain length is rarely addressed.

In this report, we have demonstrated strain measurement using SMS configuration. The novelty of our work lies in incorporation of multi point assessment. Apart from this, we also utilize intensity interrogation system in lieu of wavelength interrogation which makes the overall sensing unit quite portable as well as less expensive. All results hitherto are analyzed and interpreted.

## 2. Theory:

The SMS configuration is shown schematically in the figures below.

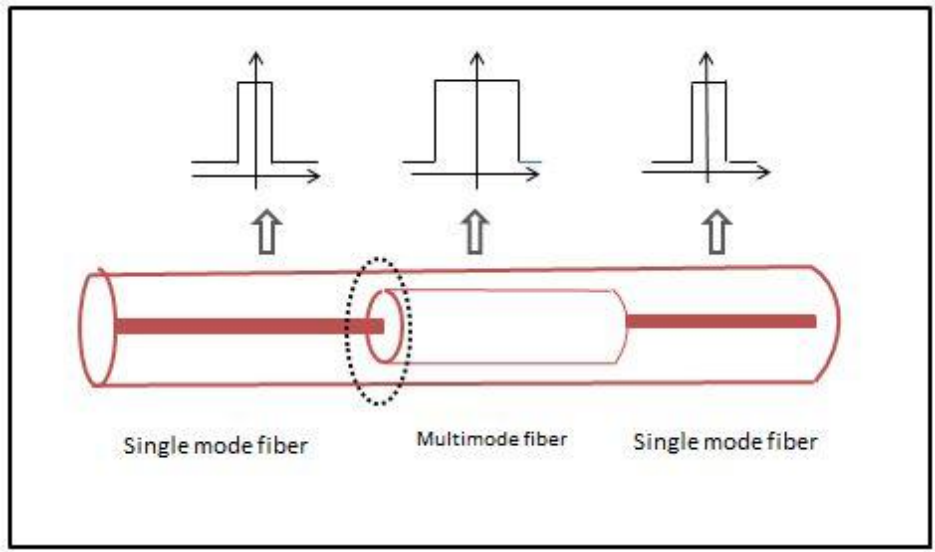

Figure 1. SMS design

As shown in Fig 1, we splice one single mode fiber with one end of a multimode fiber. The other end is again spliced with identical single mode fiber. Owing to a perfect alignment with the core axis, we can assume generation of circularly symmetric modes in the multimode fiber. To be more specific, we have two ports in this configuration. One is lead in and the other is lead out. In the lead in side, single mode fiber is conjoined with mid multimode fiber (MMF). Light enters through lead in and reaches the junction of single mode-multimode. Since the entry level is single mode fiber. As such, there will be fundamental mode dominance. In other words, there is core-guided mode. When light steps from single mode fiber to multimode fiber, the fundamental mode dominance vanishes. Along with core modes, there arise other guided modes such as cladding modes. All these excited modes have different propagation constants. Therefore, when these modes make their exit through lead out port, multimode interference occurs. This is due to the fact that MMF is followed by second SMF. As stated earlier, these different propagation constant valued modes give rise to phase difference. Let us suppose that we change temperature or pressure in immediate neighborhood of SMS. This will lead to alteration in propagation constants and eventual phase difference. The ultimate output will be reflected in the coupled light from the mid MMF to SMF. Accordingly, there will be different transmittance. By measuring transmittance or output power, one can directly have an assessment of the physical variable.

Various excited modes in the MMF possess different propagation constants along length $L$ of MMF. In the lead out junction, i.e., at the second splice, they get coupled by making exit through the lead-out SMF. The output is expressed as

$$P = \left| \int_0^{2\pi}\int_0^\infty \psi_s^* \psi(z=L) r\, dr\, d\phi \right|^2$$

$$= \left| A_0^2 + A_1^2 \exp[i(\beta_0 - \beta_1)L] + A_2^2 \exp[i(\beta_0 - \beta_2)L] + \ldots \right|^2$$

In the above expression, $\psi_s$ and $\psi$ are single mode fiber wave function and MMF wave function. $A_0$, $A_1$ are coefficients. Similarly, $\beta_0$ and $\beta_1$ are propagation constants with respect to fundamental and higher order modes.

It is apparent that above equation is adequate to establish the dependence of output power on relative phase difference between fundamental and higher order modes due to MMF. Besides, external perturbations play another deciding factor in bringing out these relative phase difference. Exploiting this property, we implement a simple SMS configuration to measure multi-point strain.

3. Preparation of SMS fiber structure:

The SMS fiber structures are prepared through fusion splicing. It is prepared by splicing a SMF with one end of a MMF. The other remaining end is again spliced with another identical SMF. SMF possesses core diameter ~5 μm along with a cladding diameter of ~125 μm. Similarly, the MMF has 105 and 125 μm as core and cladding diameters. In the fabrication of SMS scheme, we chose a length of 60 mm for the mid MMF which spliced with SMFs on either side.

4. Experimental Design:

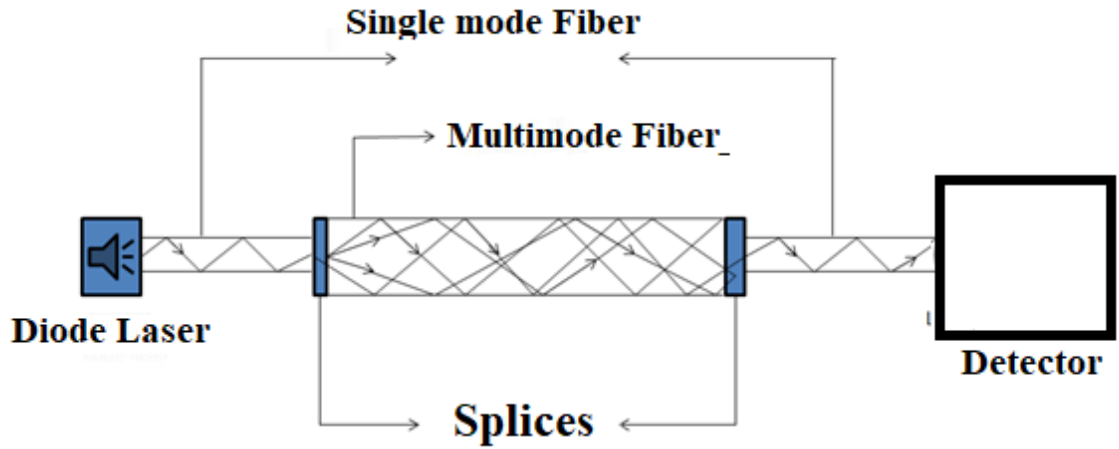

(a)

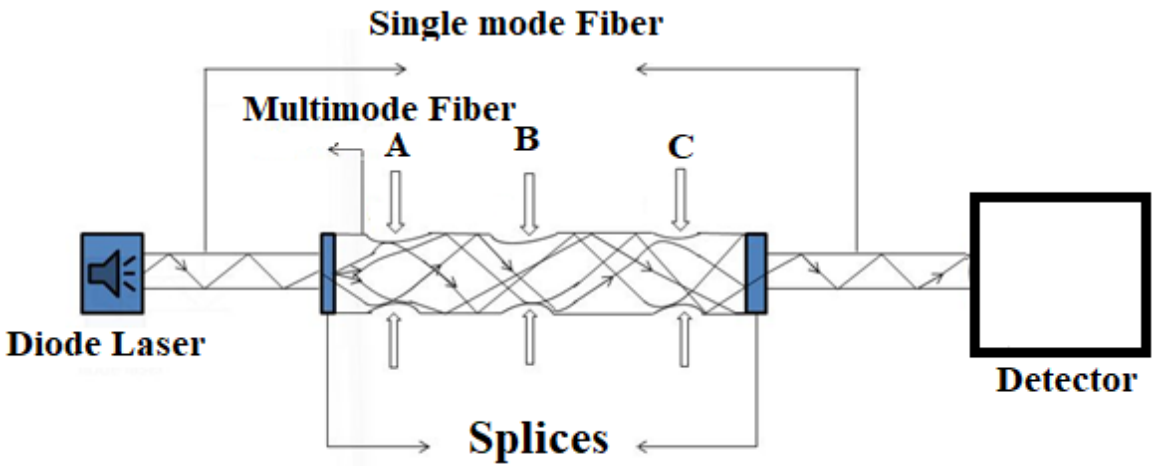

(b)

Figure 2. (a) Schematic of experimental set-up without application of strain. (b) Same as in 'a' but with strain applied at multipoint.

We have used a diode laser, SMS configuration. The strain measurement points are at three defined locations A, B and C. In order to apply strain, we used a slider translational stage arrangement. These three points are subjected to equal levels of microstrain. Accordingly, they are augmented in equal intervals. To analyze output in terms of transmittance, we have utilized an optical detector (Thorlabs, USA). In order to apply strain in SMS structure, the unit is firmly held between two clamps. Then, one of the ends is mounted on a translational stage setting the microstrain levels $\mu\varepsilon$. Accordingly; the translational stage is moved along the points, there by exerting identical levels of strain. These three points are equidistant from each other. Point B is situated in between point A and point C. While applying strain, utmost care is taken that it does not percolate to splice protector. Or else, it will either lead to damage of the SMS configuration or may lead to erroneous output. All the measurements are carried out on a level surface so that no external strain is developed except the applied one. In order to have more insight into this measurement, we exe

5. **Results and discussions**

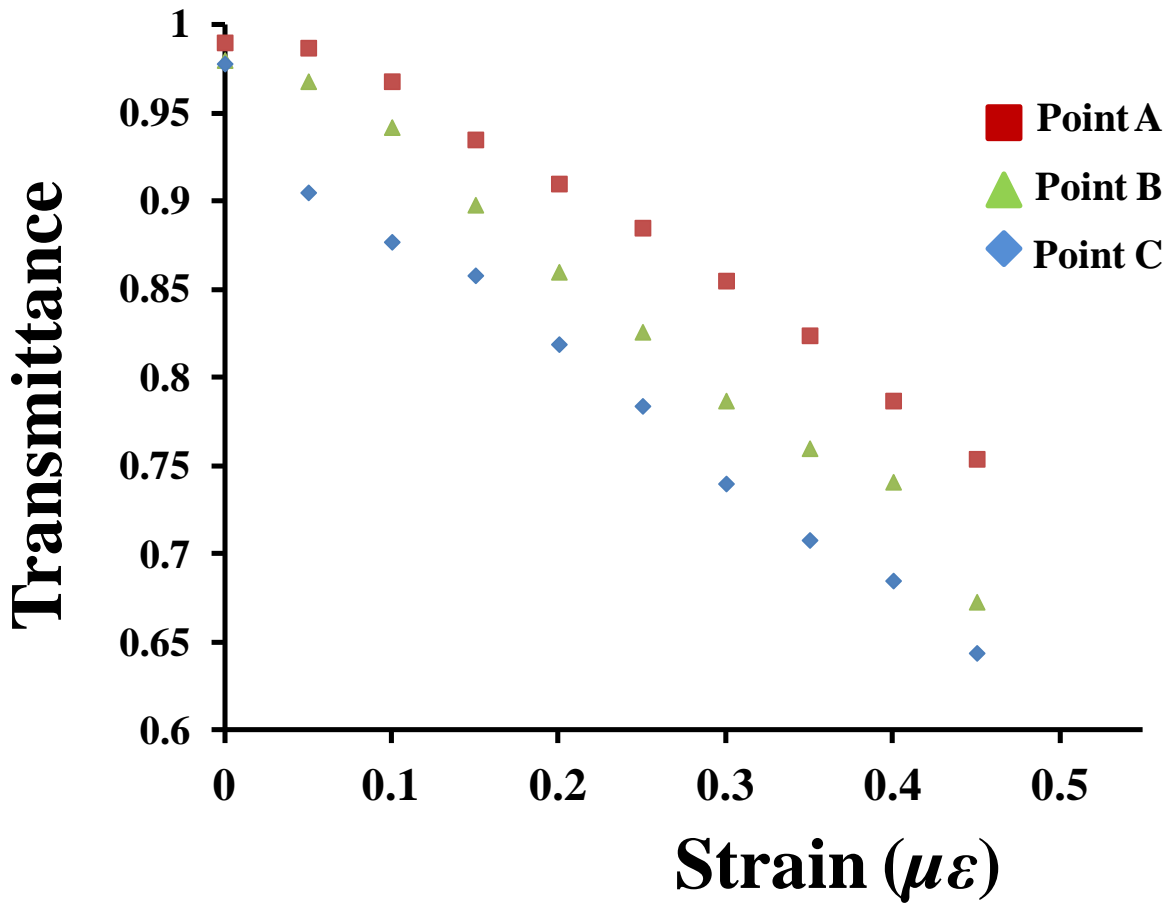

Figure 3. Plot between transmittance vs. strain. Transmittance is provided as arbitrary units.

Figure 3 depicts the pattern of transmission corresponding to the fiber optic set-up. As stated in the experimental design section, we have taken three points in the mid single mode fiber section at equal intervals. We defined the points as A, B and C. Here, point A is nearest to the spliced portion of MMF-SMF section. In other words, it is just in proximity in the input irradiance. Point B is located in the mid portion of the fiber. Likewise, point C is the farthest point from the input section, however, at the same time, it is the nearest to the receiving side (MMF-SMF) section.

Corresponding to these points, we observe the decline profile while subjecting them to equal level of micro-strain. The transmittance trend was as per anticipation. When there is no application of strain, the transmittance switches back to maximum intensity. As the strains are exerted, we see variation in the output intensity. Precisely, the variation seems to tally well with positions of the strain points.

First we consider point A. As it is in immediate neighborhood to the input section, hence, we observe highest transmittance at zero level of strain. However, as we slowly increase magnitudes of strain, we see an approximate linear decline in the observed transmittance. However, point B, we see further decline in the intensity. As it is comparatively at a higher separation as compared to A, as such the diminution in levels of intensity is quite normal. However, the attained values appear to be less linear. Lastly, after subjecting to strain, we noted response for point C. In this case, the decrease in intensity level is highest as compared to point A and B.

The trend of intensity distribution is found to follow with the locations of the measuring points where we have applied strain. As B is between point A and C, the transmittance trend also exhibits the identical trend. All derived intensity values corresponding to point B is sandwiched between values of A and C. Since C is farthest, it yields lowest transmittance profile. Apart from that, the intensity magnitudes corresponding to point B seems to be more non-linear as compared to the other points. Accordingly, when we tried exploring the linearity of the values corresponding to point C, it exhibits a linear regression coefficient of ~0.993.

If we look at the underlying process which contributes to this higher linearity, the multimode interference (MMI) can be taken into account as one of the plausible reason [2-7]. As we are considering a SMS hetero-core fiber structure, hence, MMI effect leads to self imaging. Since point C, although the farthest point with respect to the input section, is just near the output section, as such it directly falls in the region of MMI effect, giving rise to a very good response with increase in strain levels. Likewise, point A registers non-linear output as regards to change in strain levels. Actually, input irradiance goes through the SMF section and then it enters the MMF; thus exciting core-cladding modes. As point A is nearest, we attain a higher intensity. As the light progresses point B where strain is applied, the output loses its linearity. This transition to non-linear response can be interpreted as the intermediate stage of MMI effect. Moreover, the competition between the core and cladding modes can be reckoned to be contributing to this. Over all, the whole set-up is yielding a considerable response. In order to see whether the findings are reproducible, we reverse input SMF-MMF and output MMF-SMF section and carry out the results with inclusion of three identical points accompanied by similar magnitudes of strain. All the results are found to be same as the previous set-up. Apart from that, we replicate our observation several times and we attain similar trends that establish the repeatability of our set-up. As compared to long period gratings or fiber brag gratings, this proposed scheme works excellently without being perturbed by temperature change. Owing to absence of spectrometer or optical spectrum analyzer, this scheme proves to be more cost effective with inclusion of cost-effective interrogation of intensity modulation.

## 6. Conclusion

In summary, we have reported a highly functional hetero-core spliced fiber optic set-up for measuring strain on multipoint basis. The sensing system is found to be endowed with good repeatability as well as high sensitivity. We find an excellent response so far as multipoint strain assessment is concerned. It is envisioned that this sensing set-up will prove beneficial for detecting failure and crack propagation where the modulated output in the form of strain variation can provide vital information towards it.